\documentclass[pre,reprint,twocolumn,showpacs,floatfix,amsmath,amssymb]{revtex4-1}
\usepackage{xcolor}

\usepackage{amsmath, amssymb, amsfonts, amsthm}
\usepackage{verbatim}
\usepackage{hyperref}
\usepackage{float} 
\usepackage[caption=false]{subfig}
\usepackage{graphicx,color}
\usepackage{silence}
\newcommand{\figureOne}{
\begin{figure}[htb!]
\includegraphics[width=.45\textwidth]{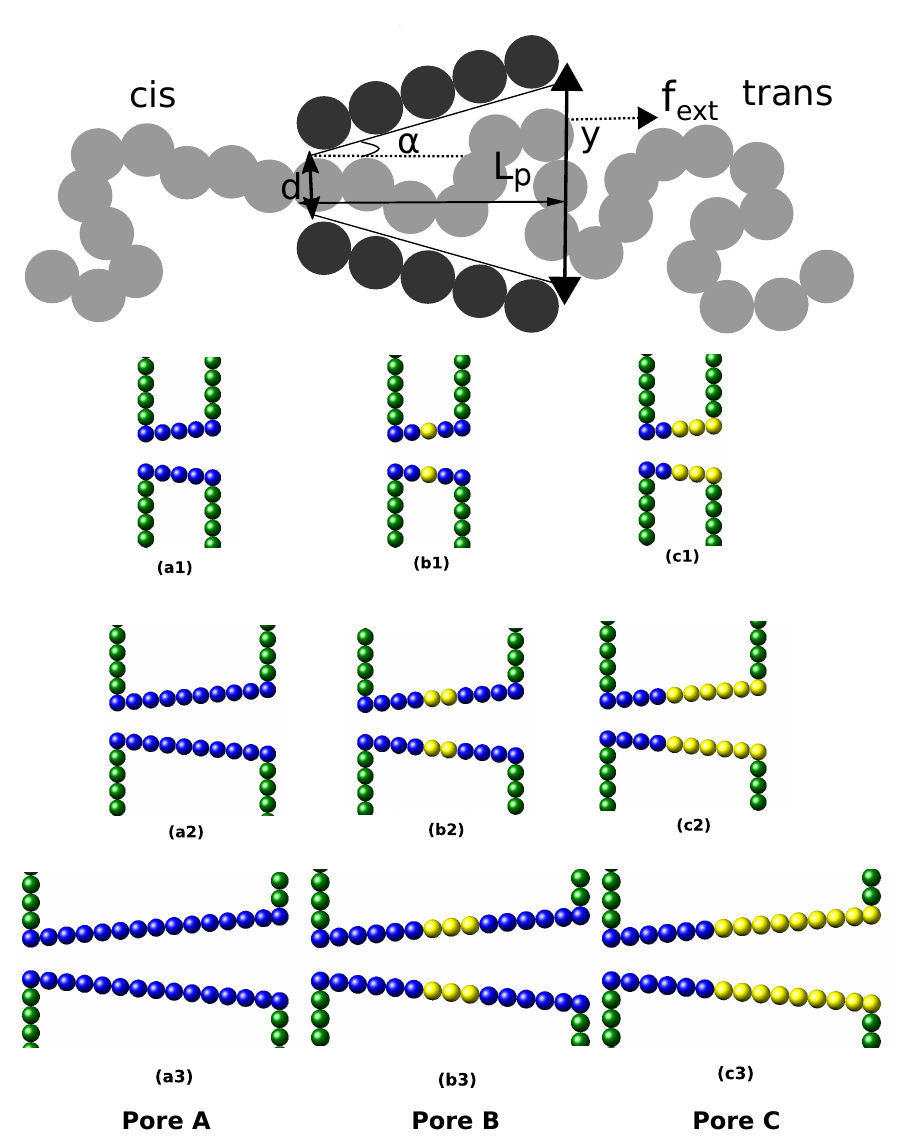}

	\caption{Top row shows a schematic diagram of pore-polymer set up
	for translocation in two-dimensions. The pore is 5 beads long with a half apex angle $\alpha$. The pore diameter on the cis end is
	$d=2.25\sigma$, whereas it is $y = d + 2L_p \tan \alpha$ on the
	trans end. The flexible polymer is made up of beads with diameter
	$\sigma=1$ connected with springs. The subsequent rows show
	patterned pores with different pore lengths, $L_{p}=$ 5, 10, and 15,
	respectively. The blue beads are attractive while the yellow and green beads
	are repulsive in interaction with polymer chain.  }

\label{fig:1}
\end{figure}}

\newcommand{\figTwo}{
\begin{figure}[t]
\includegraphics[width=0.45\textwidth]{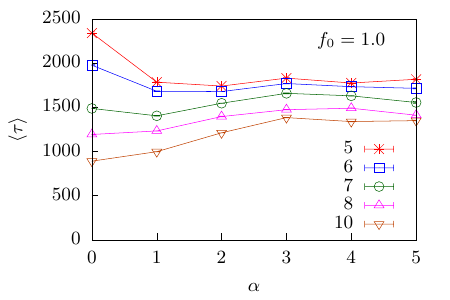}

	\caption{Plots showing the variation of $\tau$ vs. $\alpha$
	behavior, with the change of pore-length, $L_{p}$. Here,
	the polymer is flexible with length $N$ of 128 beads, and pore-length
	varies from 5$\sigma$ to 10$\sigma$. }\label{fig:2}

\end{figure}}

\newcommand{\figThree}{
\begin{figure}[htb!]
\includegraphics[]{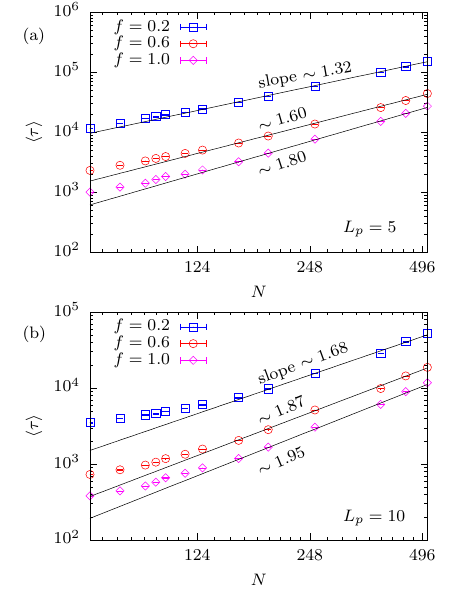}

	\caption{$\langle \tau \rangle$ as a function of $N$ (in log-log
	scale) for three different driving forces when the polymer is
	translocating from an extended cylindrical pore (i.e., $\alpha = 0$)
	of length (a) $L_p = 5\sigma$, and (b) $L_p = 10\sigma$. The slopes
	mentioned in the plot are the values of the translocation exponent,
	$\beta$, defined by $\langle \tau \rangle \sim N^{\beta}$, for
	different driving forces.} \label{fig:3}

\end{figure}
}

\newcommand{\figFour}{
\begin{figure*}[htb!]
\centering
\includegraphics[]{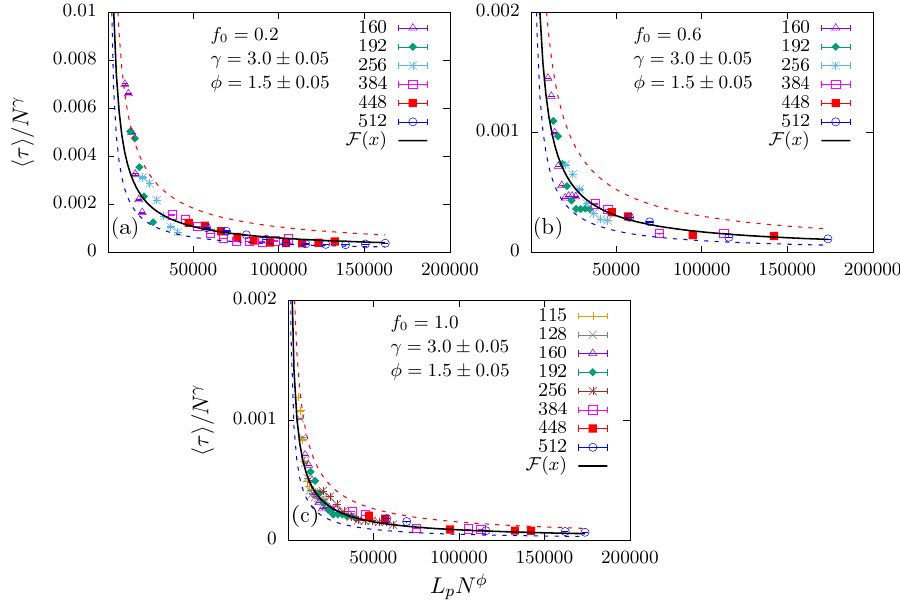}

	\caption{$\langle \tau \rangle / N^{\gamma}$ as a function of $L_p
	N^{\phi}$ for an extended pore ($\alpha = 0$) for three different
	driving forces (a) 0.2, (b) 0.6, and (c) 1.0. The lines are the
	scaling function $\mathcal{F}(x)\sim 1/x^p$ with exponent $p=0.85$.The dashed line above (below) is for exponent value $p = 0.80 \ (p = 0.90)$  }
	\label{fig:4}

\end{figure*}
}

\newcommand{\figFive}{
\begin{figure*}[htb!]
\centering
\includegraphics[width=\textwidth]{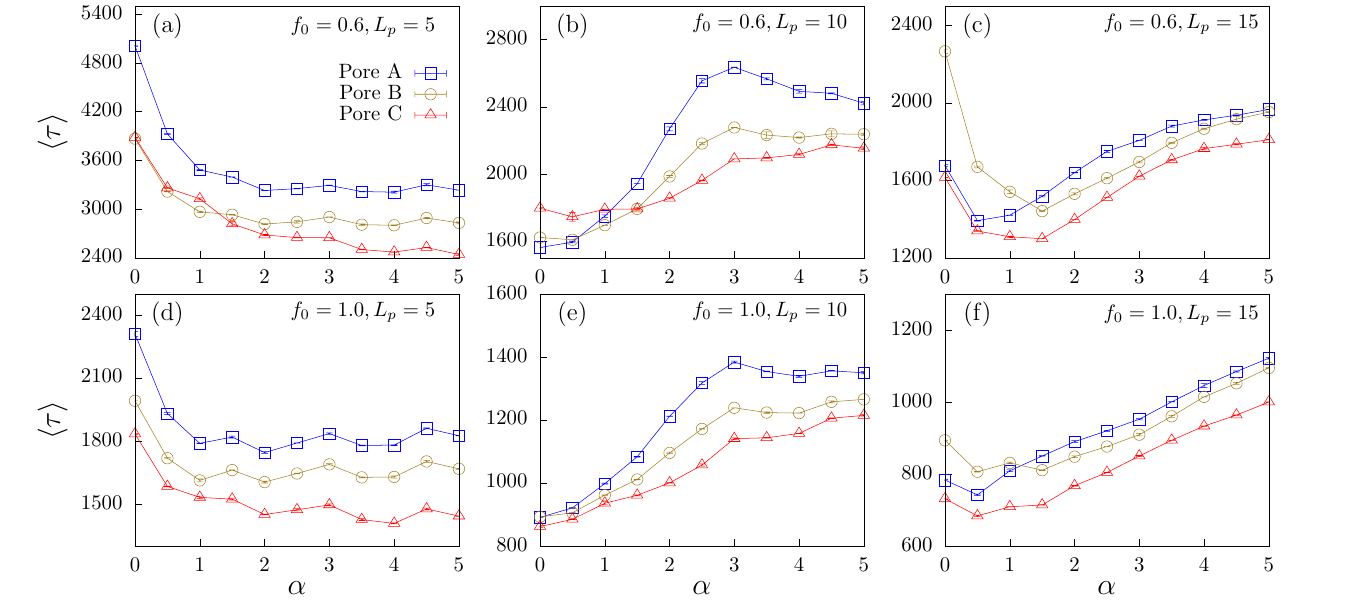}

	\caption{Comparison of $\tau$ vs. $\alpha$ for three different pore
	sizes $L_{p}=$ 5, 10 and 15 for two different forces $f_{0}= 0.6$
	and 1.0. The length $N$ of the flexible polymer chain is 128
	beads.}\label{fig:5}

\end{figure*}}

\newcommand{\figSix}{
\begin{figure*}[htb!]
\centering
\includegraphics[width=0.8\textwidth]{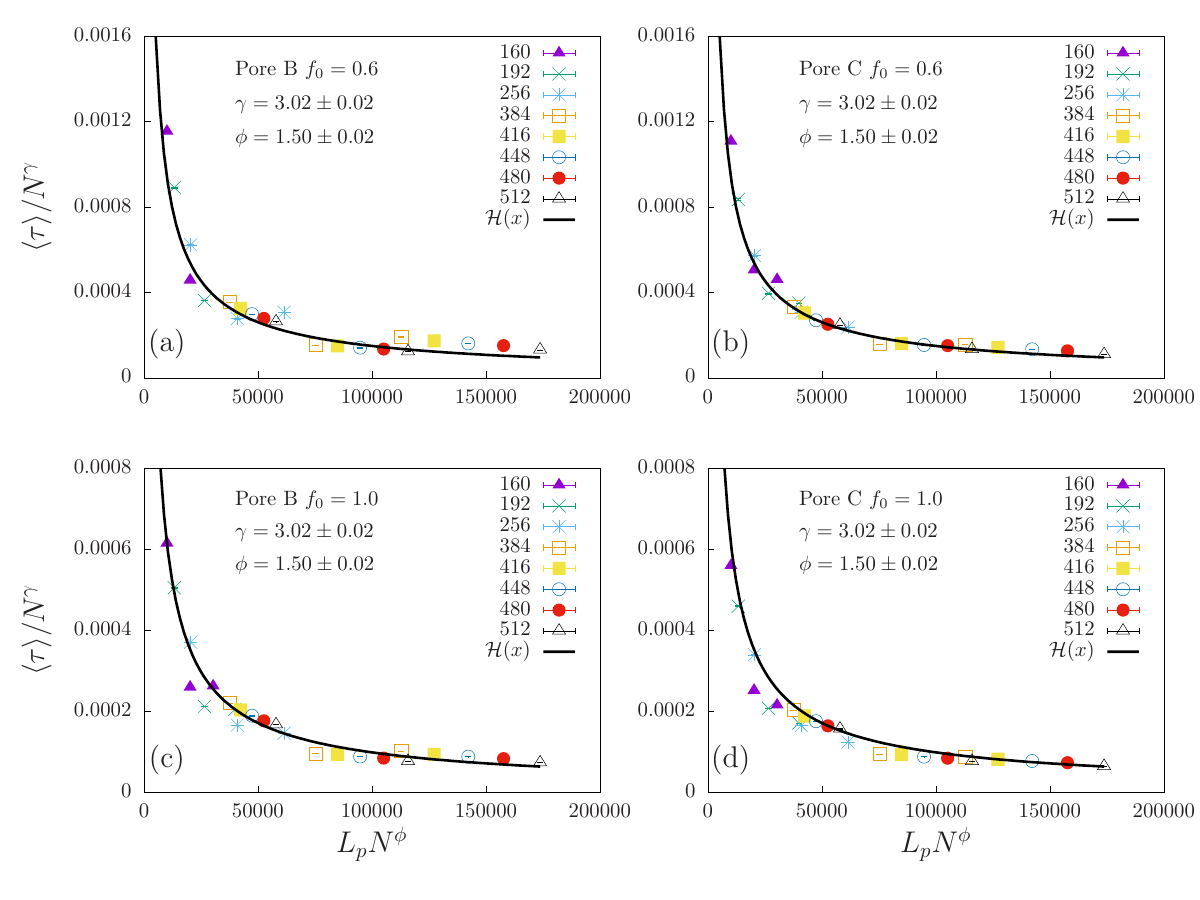}

	\caption{$\langle \tau \rangle/N^{\gamma}$ as a function of
	$L_pN^\phi$ for patterned pores: Pore B and Pore C for $\alpha = 0$
	and three different pore lengths $L_p =5$, 10 and 15 and various
	chain lengths $N$ as indicated in the plots. Plots (a), (b) are for
	force $f_0 = 0.6$ and plots (c), (d) are for force $f_0 = 1.0$,
	respectively. The solid line represents the scaling function of the form
	${\mathcal H}(x) \sim 1/x^{p}$ with exponent $p=0.8$.
	}\label{fig:6}

\end{figure*}
}

\newcommand{\figSeven}{
\begin{figure*}[htb!]
\centering
\includegraphics[width=0.85\textwidth]{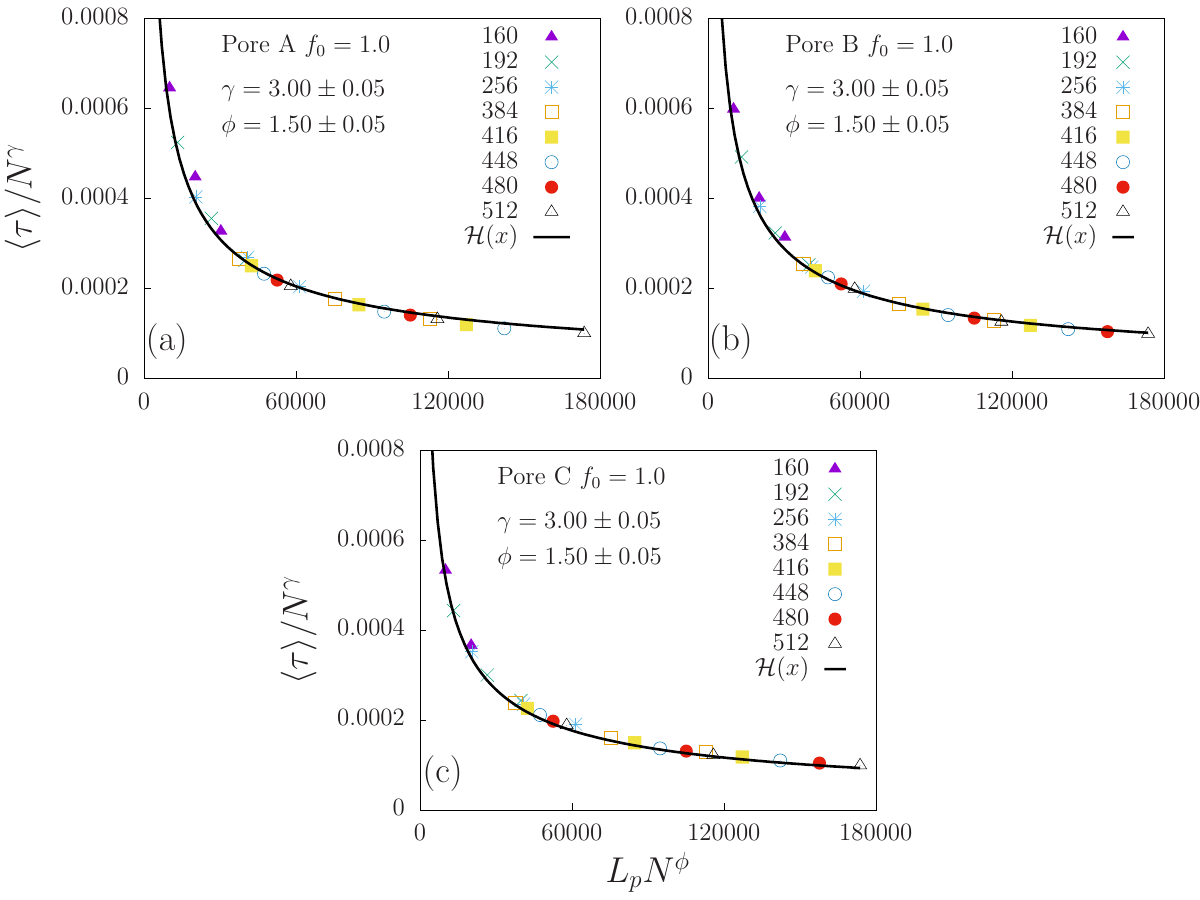}

	\caption{$\langle \tau \rangle/N^{\gamma}$ as a function of
	$L_pN^\phi$ for patterned pores (a) Pore A, (b) Pore B and (c) Pore
	C for $\alpha = 3^{\circ}$ and $f_0=1.0$. Three different pore
	lengths used are $L_p =5$, 10 and 15 and various chain lengths $N$
	as indicated in the plots. The solid line indicates the scaling function of
	the form ${\mathcal H}(x) \sim 1/x^{p}$ with exponent
	$p=0.6$.}\label{fig:7}

\end{figure*}
}

\begin{document}

\title{
Polymer translocation through extended patterned pores in two dimensions: Scaling of the total translocation time}

\author{Andri Sharma$^{1,2}$}
\email{sharmaandri1@gmail.com}
\author{Abhishek Chaudhuri$^1$}
\email{abhishek@iisermohali.ac.in}
\author{Rajeev Kapri$^1$}
\email{rkapri@iisermohali.ac.in}

 \affiliation{$^1$ Department of Physical Sciences, Indian Institute of
 Science Education and Research (IISER) Mohali, Sector 81, S. A. S.
 Nagar 140306, Mohali, Punjab, India}

 \affiliation{$^2$ National Centre for Biological Sciences-TIFR, Bellary
 Road, Bengaluru 560065}

% \date{}% It is always \today, today,

\begin{abstract}

	We study the translocation of a flexible polymer through extended
	patterned pores using molecular dynamics (MD) simulations. We
	consider cylindrical and conical pore geometries that can be
	controlled by the angle of the pore apex $\alpha$. We obtained the
	average translocation time $\langle \tau \rangle$ for various chain
	lengths $N$ and the length of the pores $L_p$ for various values
	$\alpha$ and found that $\langle \tau \rangle$ scales as $\langle
	\tau \rangle \sim N^\gamma \mathcal{F}\left( L_p N^\phi \right)$
	with exponents $\gamma = 3.00\pm0.05$ and $\phi = 1.50\pm0.05$ for
	both patterned and unpatterned pores, respectively.

\end{abstract}

\maketitle

\section{Introduction}\label{intro}

Polymer and protein translocation through narrow pores is a fundamental
process in biology and an emerging paradigm for nanoscale transport in
technology.  Translocation of polymer or proteins is an ubiquitous
process inside biological cells and is among one of the important
processes that propel the cell
functionality~\cite{Albe_2002_book,van2007motor}. In the past two and
half decades, polymer translocation has gained considerable attention
due to its technological applications such as rapid DNA sequencing, gene
therapy and controlled drug delivery~\cite{salman2001pnas,ying2022nn}.
Recent advances in nanofabrication have enabled stable synthetic
nanopores that can operate under high voltages, elevated temperatures,
and various solvent conditions. These pores have become controllable
experimental systems to explore the physics of single‐molecule
transport.  A single strand of a biopolymer such as DNA or RNA can be
electrophoretically driven through such nanopores in a controlled
manner. The ion current flowing through the nanopore gets blocked when
the polymer is inside the pore which gives the signature of the polymer
sequence~\cite{kasianowicz1996characterization}.        

The translocation of polymer through nanopores depends on various
parameters, such as, the length of the polymer chain, its  structure,
shape and the size of the pore and pore-polymer
interactions~\cite{Muthukumar2007-yc,muthukumar2}. Consequently, polymer
translocation is a complex, system-dependent nonequilibrium process
whose understanding requires both theoretical modelling and
computational simulations.  Various coarse-grained models have been used
to understand the translocation
phenomenon~\cite{sung1996polymer,muthukumar2,palyulin2014polymer,rowghanian2011force,sakaue2007nonequilibrium,kantor2004anomalous,dubbeldam2007driven,ikonen2012unifying}.
For example, the translocation process has been studied as a
one-dimensional barrier crossing problem of the translocation coordinate
through a slit (i.e., a pore of unit length). In initial studies, it was
assumed that the translocation time is long enough to ensure the
equilibration of the polymer conformations at every stage of the
translocation
process~\cite{sung1996polymer,muthukumar1999polymer,muthukumar2}. Later,
a majority of experimental studies with synthetic nanopores showed that
the time taken by the polymer to pass the nanopore is much less than its
relaxation time, indicating that translocation is a non-equilibrium
process. The tension propagation (TP) theory proposed originally by
Sakaue~\cite{sakaue2007nonequilibrium} for an infinite chain and
subsequently modified by Ikonen et
al.~\cite{ikonen2012unifying,ikonen2012influence,sarabadani2014iso,sarabadani2018theory}
and Dubbeldam et al.~\cite{dubbeldamPRE2012} for finite chains has
proven to be successful in explaining the non-equilibrium aspect of
driven polymer translocation~\cite{chen2021dynamics}. The most important
quantity of interest in polymer translocation studies is the dependence
of the average translocation time, $\langle \tau \rangle$, on the chain
length $N$. Numerical simulations and experiments have indicated that
$\langle \tau \rangle \sim N^{\beta}$. Different values of $\beta$ have
been observed in different studies, suggesting that $\langle \tau
\rangle$ depends on various physical
parameters~\cite{milchev2011single}. For a self-avoiding flexible chain,
$\langle \tau \rangle$ is known to scale as $\langle \tau \rangle \sim
N^{1+\nu}$, here $\nu=3/4$ in two
dimensions~\cite{chuang2001anomalous,luo2007influence,sarabadani2018theory},
and $\nu \approx 0.59$ in three dimensions~\cite{
	doi1988theory,chuang2001anomalous,kantor2004anomalous}.

While early theoretical work considered a very short or unit‐length
pore, most realistic nanopores are extended and of finite length, often with diameters of tens of monomer long. In such pores, the details of the
pore–polymer interaction and spatial patterning along the pore axis
strongly influence the
dynamics~\cite{storm2003fabrication,wanunu2007chemically,chen2004probing,lan2011pressure,liu2012voltage}.
%Experimental studies on polymer translocation involve pores of
%finiteclengths in which pore-polymer interactions play a dominant role.
There have been theoretical and computational studies on finite-length
pores where the shape of the extended pore is cylindrical, with
different patterned pore-polymer
interactions~\cite{luo2007influence,luo2008translocation,cohen2012stochastic,cohen2012translocation,katkar2018role,kumar2018sequencing},
or
conical~\cite{nikoofard2013directed,nikoofard2014accuracy,nikoofard2015flexible,sharma2022driven,Sharma_2024,upadhyay2024epje,upadhyay2024jpcm}.
For cylindrical pores, it was estimated that for a fixed external drive
$\langle \tau \rangle$ depends on the patterning of the pores. It was
found that $\langle \tau \rangle$ is the maximum for a fully attractive
pore and is the minimum for a pore having a repulsive
exit~\cite{cohen2012translocation,kumar2018sequencing}. It was argued
that the pattern stickiness of extended pores can be exploited to
predict heteropolymer sequences with different bending rigidities to a
high degree of accuracy by translocating them through multiple
pores~\cite{kumar2018sequencing}. For conical pores, we have one more
parameter--the cone apex angle, which we represent by $\alpha$. For
$\alpha = 0$, we have a cylindrical pore. For a conical pore with
attractive pore-polymer interactions, $\langle \tau \rangle$ was found
to be a nonmonotonic function of $\alpha$ for lower driving forces,
which becomes monotonic for higher
forces~\cite{sharma2022driven,Sharma_2024}. 

The length of the extended pore, $L_p$, introduces a second independent
length scale in addition to the polymer contour length $N$.
Understanding how these two compete under different interaction patterns
is essential for controlling transport through synthetic or biological
nanopores.  Although there have been many scaling studies of $\langle
\tau \rangle$ on the length of the polymer $N$, not much is discussed in
the literature on the dependence of $\langle \tau \rangle$ on $L_p$.  We
try to address this in this paper.   

In this work we use two-dimensional Langevin dynamics simulations to
study the driven translocation of a flexible polymer through extended
conical pores having different attractive–repulsive patterns. The
two-dimensional approach enables long-chain statistics and extensive
averaging, while retaining the essential physics of polymer stretching
and confinement.  We extend our work on conical
pores~\cite{sharma2022driven,Sharma_2024} to study the effect of
different pore patterns on the driven-translocation process. The
patterning is created along the pore length by having various sections
with attractive and repulsive interactions.  The entrance of the pores
is always attractive, so that the polymer can enter the pore with ease.
The exit from the pore can either be attractive or repulsive.  The
presence of an external drive along the channel often facilitates the
translocation process. We present our study on translocation behavior by
varying the pore length $L_p$ and the polymer length $N$.  The key
questions we address are: (i) how does the mean translocation time depend
simultaneously on $N$ and $L_p$, and (ii) whether distinct pore patterns
produce universal or pattern-specific scaling behavior.

The rest of this paper is organized as follows.  In Sec.~\ref{model}, we
define our model and simulation details. In Sec.~\ref{results}, we study
the effect of the pore length, $L_p$, and chain length, $N$, on $\langle
\tau \rangle$. We also discuss the effect of pattered pores on $\langle
\tau \rangle$ and establish scaling relations between $\langle \tau
\rangle$, $L_p$ and the polymer length $N$ for extended pores. In
Sec.~\ref{discussion}, we conclude by providing a discussion of the
obtained scaling exponents.

\figureOne

\section{Model}\label{model}

All simulations are performed in two dimensions (2D). The reduced
dimensionality allows extensive statistical sampling and direct
comparison with earlier 2D translocation
studies~\cite{chuang2001anomalous,luo2007influence}. The essential
physics of polymer stretching and confinement within narrow pores
remains captured in 2D while significantly lowering computational cost.

The polymer is
modeled as a bead-spring chain~\cite{grest1986molecular}, where the
consecutive beads with diameter $\sigma$ are connected with a spring of
spring constant $K$ interacting via a harmonic potential given by:
\begin{equation} \label{eq:Ubond}
    U_{\textrm{bond}}=\frac{1}{2}K(r-r_0)^{2},
\end{equation}
where $r_{0}=1.0\sigma$ is the equilibrium bond length. 
The non-bonded polymer beads
interact via the Weeks-Chandler-Andersen (WCA) potential $U_{\textrm{nb}}
(r)$ of the form: 
\begin{equation} \label{eq:Unonbonded}
U_{\textrm{nb}}(r)=
  \begin{cases}
  4 \epsilon \Bigg[
	  \left(\frac{\sigma}{r}\right)^{12}-\left(\frac{\sigma}{r}\right)^{6}
	  \Bigg]+\epsilon,& r < r_{\rm min}\\ 
  0,& r \geq r_{\rm min}, 
  \end{cases}
\end{equation}
where $\epsilon$ is the strength of the potential. The cutoff distance
$r_{\rm min}=2^{1/6}\sigma$ is set at the minimum of the potential. 
Unless otherwise stated, we set $\epsilon=1.0$ and use the same energy scale for all interactions.

The pores are static and are constructed with coarse-grained beads of size
$\sigma$. We study three different types of extended pores as considered
in Ref.~\cite{kumar2018sequencing}:

\begin{enumerate}
    \item[(1)] Pore A is an attractive pore. Polymer beads interact with
		pore beads via the attractive Lennard-Jones (LJ) potential,
		$U_{\textrm{pp}}(r)$:
    \begin{equation} \label{eq:Uporepolymer}
    U_{\textrm{pp}}(r)=
    \begin{cases}
        4 \epsilon_p \Bigg[
			\left(\frac{\sigma}{r}\right)^{12}-\left(\frac{\sigma}{r}\right)^{6}
			\Bigg],& r < r_{c}\\ 
        0,& \text{otherwise},
    \end{cases}
    \end{equation}
    where $\epsilon_p=1.5\epsilon$ is the attractive strength of the pore and $r_c
		= 2.5\sigma$ is the cut-off distance.

    \item [(2)] Pore B has attractive entrances and exits separated by a
		repulsive section. The polymer beads interact with attractive
		beads of the pore via the LJ potential and with beads of the
		repulsive section via the WCA potential.  

    \item [(3)] Pore C has an attractive entrance and a repulsive exit.
		The polymer beads interact with the beads of the attractive
		section of the pore via the LJ potential and the beads of the
		repulsive section via the WCA potential.  
\end{enumerate}
The entrance of all pores is kept attractive to ensure capture of the polymer, while the exit region determines the ease of release. The total pore pattern is scaled proportionally when the pore length $L_p$ is varied so that the relative fraction of attractive and repulsive segments remains constant.

In addition to the above potentials, an external pulling force,
$\vec{f}_{\textrm{ext}}$, which mimics an electrophoretic drive, is applied inside the
pore in the positive $x$-direction to ensure that the polymer
translocates through the pore from the \emph{cis} to the \emph{trans}
side. The force magnitude varies along the pore extension as
\begin{equation}
f_{\textrm{ext}}(x) = \frac{ f_0 d}{(d + 2x \tan \alpha)},
\label{eq:fext}
\end{equation}
where $f_{0}$ is the driving force at the pore entrance of width $d$.
For a conical pore ($\alpha>0$), this form maintains a constant total integrated force along the pore length, causing the local field to decrease gradually from the entrance to the exit. For $\alpha=0^\circ$, the same equation reduces to a constant field in a cylindrical pore, as expected.
The schematic diagram of our coarse-grained model is
shown in Fig.~\ref{fig:1} (top row). The subsequent rows show the
schematic diagram of three different pore patterns that we study in
this paper. The second row shows three different patterned pores for
length $L_p = 5$: 
Fig.~\ref{fig:1}(a1) corresponds to a fully attractive pore (Pore~A), 
Fig.~\ref{fig:1}(b1) shows a pore with attractive entrances and exits but a repulsive middle segment (Pore~B), and 
Fig.~\ref{fig:1}(c1) shows a pore with an attractive entrance and a repulsive exit (Pore~C). 
In the third and fourth rows, we show pores for lengths $L_p=10$ and $L_p=15$,
respectively. 

The dynamics of the polymer is governed by the stochastic Langevin
equation. The equation of motion for the position of the $i$th monomer is
\begin{equation}
\label{eq:LD}
    m \ddot{\vec{r_i}} = -\eta\dot{\vec{r_i}} - \nabla_i (U_{\textrm{bond}} +
	U_{\textrm{nb}} + U_{\textrm{pp}}) + \vec{f}_{\textrm{ext}} + \vec{\zeta_i},
\end{equation}
where $m$ and $\vec{r_i}$ are the mass and position of the $i$th bead, respectively,
$\eta$ is the friction coefficient of the fluid, and $\vec{\zeta_i}$ is a random
Gaussian noise obeying the fluctuation–dissipation relation 
$\langle \zeta_{i\mu}(t)\zeta_{j\nu}(t^{\prime}) \rangle = 2
 k_{B}T\eta\delta_{ij}\delta_{\mu\nu}\delta(t-t^{\prime})$.

In our model, $\epsilon$, $\sigma$ and $m$ set the units of energy,
length, and mass, respectively. This sets the unit of time as $(m
\sigma^2/\epsilon)^{1/2}$. The force value is set in units of
$\epsilon/\sigma$ and the spring constant in units of
$\epsilon/\sigma^2$. Using this system of units, the dimensionless parameters $k_B
T = 1$, $\eta = 1$, $d = 2.25$, and $K=500$ are chosen in the
simulations. 
The choices of $\eta$ and $d$ follow standard coarse-grained studies~\cite{luo2008translocation,cohen2012translocation}, ensuring realistic frictional damping and pore geometry comparable with experimental nanopores.

For every simulation data point, the results are averaged
over 1500--2000 successful translocation events. Error bars representing one standard error of the mean are shown in all plots.
Molecular-dynamics trajectories are
obtained using LAMMPS~\cite{thompson2022lammps} software.

\section{Result and Discussion}\label{results}

\subsection{Effect of pore length} 

\figTwo
\figThree
\figFour

The length of the pores, $L_{p}$, plays an important role in the
translocation of a polymer through a membrane channel. To study how the
translocation time depends on the length of the pores, we consider a
polymer of length $N=128$ and translocate it through pores of varying
lengths $L_p \in [5\sigma, 10\sigma]$. (Error bars in all figures denote one standard error based on 1500--2000 successful events.)
For each pore length, we also
vary the angle of the cone apex, $\alpha$, in the range $\alpha \in
[0,5]$ degrees. The magnitude of the external force is kept constant at
$f_{0} = 1$. The pore-polymer interaction is attractive in nature and is
modeled using Eq.\eqref{eq:Uporepolymer}. In Fig.~\ref{fig:2}, we have
plotted the average translocation time $\langle \tau \rangle$ as a
function of $\alpha$ for various $L_p$.

Figure~\ref{fig:2} shows that (i) for cylindrical pores
(\(\alpha=0^\circ\)) the mean translocation time \(\langle\tau\rangle\)
decreases strongly with pore length \(L_p\), and (ii) for conical pores
(\(\alpha>0\)) \(\langle\tau\rangle\) decreases with \(\alpha\) for
\emph{short} pores but \emph{increases} with \(\alpha\) for \emph{long}
pores. This behavior can be understood as a competition between two
effects: (a) the widening exit increases the available configuration
space for the leading segment and reduces confinement-related hindrance
near the exit (entropic relief), which facilitates the final emptying
for short pores; and (b) the imposed driving field in a cone decreases
along the pore, \(f_{\rm ext}(x)=f_0 d/(d+2x\tan\alpha)\)
(Eq.~\eqref{eq:fext}), so an extended weak-drive region develops near
the exit, making long conical pores bottleneck-limited. Because polymer
translocation is a nonequilibrium process involving tension propagation
and conformation changes along the chain, we do not attempt a
point-particle drift-time theory; instead, the finite-size scaling
analysis presented below provides the quantitative characterization of
\(\langle\tau\rangle\) across \((N,L_p,\alpha)\) and pore patterning.

\medskip

\noindent \textit{Role of attractive interactions.}
In short cones, the entropic assistance dominates and
\(\langle\tau\rangle\) decreases with \(\alpha\) before saturating once
the exit is sufficiently wide. In long cones, the weakening of the local
drive near the exit, outweighs the entropic gain; therefore, \(\langle\tau\rangle\) \emph{increases} with \(\alpha\). Attractive pore–polymer
interactions further amplify this effect by increasing residence near
the walls and promoting side-by-side monomer arrangements that impede
axial progress, again more relevant at larger \(L_p\) where the
low-drive region is longer.

\medskip

\noindent \textit{Consistency with the finite-size
scaling analysis.} The same geometry–drive competition
is reflected in the scaling form \(\langle\tau\rangle\sim
N^{\gamma}\mathcal{F}(L_p N^{\phi})\) with \(\mathcal{F}(x)\sim
x^{-p}\): for \(\alpha=0\), increasing \(L_p\) increases the number of
driven monomers and decreases \(\langle\tau\rangle\) (smaller
\(\mathcal{F}\)); for \(\alpha>0\), the weakening of the local drive
near the exit with increasing \(L_p\tan\alpha\) counteracts and
eventually reverses the entropic speed-up, consistent with the crossover
observed in Fig.~\ref{fig:2}. The measured exponents also satisfy
\(\beta\approx \gamma-(\phi p)/2\), linking the chain-length and
pore-length dependence.

\subsection{Effect of chain length $N$}
Next, we study how $\langle \tau \rangle$ varies with the chain length
$N$ for pores of fixed lengths $L_p=5\sigma$ and $10\sigma$. Let us first
consider translocation through a cylindrical pore (i.e., $\alpha = 0$).
In Figs.~\ref{fig:3}(a) and \ref{fig:3}(b), we have plotted $\langle
\tau \rangle$ as a function of $N$ for a polymer translocating through a
cylindrical pore of lengths $L_p = 5$ and 10, respectively for various
values of driving forces $f_0 = 0.2$, 0.6 and 1.0. The translocation
time exponent $\beta$ is defined as $\tau \sim N^{\beta}$. Recall that,
for a flexible polymer translocating from a slit, it was found that the
exponent $\beta < 1 + \nu$, where $\nu$ is the size
exponent~\cite{ikonen2012influence,ikonen2012unifying}. In 1- and
2-dimensions $\nu = 1.0$ and $0.75$, respectively. It is known that the
value of exponent $\beta$ depends on many factors such as chain length,
chain flexibility, driving force, friction, and pore structure. This is
also observed in our simulations. For a smaller driving force $f_0
=0.2$, we found that the numerical value of the exponent for pore length
$L_p=5$ is  $\beta = 1.32$, which is smaller than $1+\nu = 1.75$. On
increasing the pore length to $L_p = 10$, the value of the exponent
becomes $\beta = 1.68$ but still less than 1.75. On increasing the
driving force to $f_0=0.6$, the translocation exponent for $L_p = 5$ is
$\beta = 1.60 < 1.75$. However, for for pore length $L_p = 10$, the
exponent becomes $\beta = 1.87 > 1.75$. For a driving force $f_0=1.0$,
the value of the translocation exponent for pore lengths $L_p = 5$ and
10 are 1.80 and 1.95, respectively, with both being greater than 1.75 but less
than 2.

We have seen that the translocation exponent $\beta$ varies with the
length of the extended pore. We now seek to see whether the average
translocation time $\langle \tau \rangle$ for different pore lengths
$L_p$ and the polymer chain lengths $N$ follow some scaling. To check
this, we take the scaling form for $\langle \tau \rangle$:       
\begin{equation} \label{Eq:fss}
    \langle \tau \rangle \sim N^{\gamma} \mathcal{F}\left( L_p N^{\phi}
	\right),
\end{equation}
where $\gamma$ and $\phi$ are exponents, and $\mathcal{F}$ is the
scaling function. In Fig.~\ref{fig:4}, we have plotted $\langle \tau
\rangle / N^{\gamma}$ as a function of $L_p N^{\phi}$ for pore lengths
$L_p$ varying from $5$ to $10$ and chain lengths varying from $N =160$
to 512. A good data collapse is obtained for exponent values $\gamma =
3.0 \pm 0.05$ and $\phi = 1.50 \pm 0.05$ for three driven force values
($f_0=0.2$, 0.6, and 1.0) discussed in this paper. The scaling function
follows a power law, $\mathcal{F}(x) \sim 1/ x^{p}$, with a consistent
exponent $p \sim 0.85$ for each of the three forces $f_0 = 0.2$, 0.6,
and 1.0. For reader guidance, Fig.~\ref{fig:4}(a) additionally shows
dashed comparison curves generated with $p=0.80$ and $p=0.90$ to
illustrate the sensitivity of the collapse to the exponent choice. We
also discuss in Sec.~\ref{discussion} how the observed set of exponents
is consistent, within uncertainty, with the measured $\beta$ via the
relation $\beta \approx \gamma - (\phi p)/2$.

\figFive

\subsection{Effect of pore patterning} 

In this sub-section, we discuss the effect of pore patterning (shown in
Fig.~\ref{fig:1}) on $\tau$ for a polymer of length $N=128$. The
translocation of flexible and semifexible polymers through cylindrical
and conical pores has been studied in the
past~\cite{cohen2012stochastic,kumar2018sequencing,upadhyay2024epje,upadhyay2024jpcm}.
It was observed that the translocation time and its standard deviation
can be used to sequence the polymer by translocating the polymer
multiple times through different patterned pores.  Our interest in this
paper is to explore the dependence of $\tau$ on the pore length $L_p$ of
various patterned pores. We increase the pore length in such a manner
that the ratio of pore patterning remains conserved. As seen in the
second, third and fourth rows of Fig.~\ref{fig:1}, the ratio of
repulsive pore beads (yellow bead) to attractive pore beads (blue)
remains the same for three pores with pore lengths $L_{p}=5$, 10, and
15.

In Fig.~\ref{fig:5}, we have plotted the total translocation time $\tau$
for a flexible polymer of length $N=128$, as a function of the angle of
the pore apex $\alpha$ for three different pore lengths $L_p = 5$, 10,
and 15 at two different values of the driving force $f_0 = 0.6$ and 1.0.
The figure clearly shows that the behavior of $\tau$ strongly depends on
the type of pore and its length. For shorter pore lengths (e.g. $L_p =
5$), we observe that for both force values, $\tau$ decreases with
$\alpha$ and becomes saturated. This is true for the three types of
pores studied in this paper (see Figs.~\ref{fig:5}(a) and
\ref{fig:5}(d)). Upon comparing $\tau$ among these pores, we observed
that the pore with a repulsive exit (presented by the triangles) favors faster
translocation at all angles. This is followed by the pore with an
attractive entrance and exit with a repulsive bead in the middle (presented by the circles), whereas the translocation is the slowest for a fully
attractive pore (presented by the squares). For intermediate-length pores (that
is, $L_p = 10$), the translocation time $\tau$ first increases with
$\alpha$ for the force values $f_0 = 0.6$ and 1.0.  For a fully
attractive pore (see Fig.~\ref{fig:1}(a2)) it reaches a maximum at about
$\alpha \approx 3^{\circ}$ and then decreases as $\alpha$ increases
further. For pore types with repulsive beads, either in the middle (see
Fig.~\ref{fig:1}(b2)) or at the exit (see
Fig.~\ref{fig:1}(c2)), the value of $\tau$ becomes saturated around
$\alpha \approx 3^{\circ}$ (see Figs.~\ref{fig:5}(b) and
\ref{fig:5}(e)). The above-mentioned feature is more pronounced for lower force
values $f_0 = 0.6$. When the pore length increases further (e.g. $L_p =
15$), the translocation time first decreases with $\alpha$. It reaches a
minimum value at some value of $\alpha$, which is different for
different types of pores, and then $\tau$ increases as $\alpha$
increases further (see Figs.~\ref{fig:5}(c) and \ref{fig:5}(f)).

To obtain the scaling of the averaged total translocation time with pore
length $L_p$ and chain length $N$ for patterned pores, we again use the
scaling relation of the form:
\begin{equation} \label{Eq:fss1}
    \langle \tau \rangle \sim N^{\gamma} \mathcal{H}\left( L_p N^{\phi}
	\right),
\end{equation}
where $\phi$ and $\gamma$ are exponents, and $\mathcal{H}$ is the
scaling function. We observe that when $\langle \tau \rangle
/N^{\gamma}$ is plotted as a function of $L_{p}N^{\phi}$, for various
pore lengths $L_p$ and chain lengths $N$, the data collapse to a scaling
curve $\mathcal{H}(x) \sim 1/ x^{q}$. The exponents $\gamma$ and $\phi$
are remarkably robust, showing no dependence on whether the pore is
cylindrical (i.e. $\alpha = 0$) or conical ($\alpha \ge 0$) in shape.
Adjusting the proportionality constant in the scaling function
$\mathcal{H}$, we consistently obtain the exponent $q \sim 0.8$ for the
cylindrical pore and $q \sim 0.6$ for the conical pore, respectively.
Furthermore, the exponents do not vary much with $f_{0}$. The data
collapses for $\alpha =0$ and $\alpha = 3$ are shown in
Figs.~\ref{fig:6} and ~\ref{fig:7}, respectively.

\figSix
\figSeven

\section{Discussion}\label{discussion}

We investigated driven translocation of a flexible polymer through
extended pores while varying chain length $N$, pore length $L_p$, pore
geometry via cone angle $\alpha$, and axial patterning. There are three
main observations that emerge. (i) At fixed $N$, $\langle\tau\rangle$
decreases with $\alpha$ for short pores but increases with $\alpha$ for
long pores (Fig.~2), revealing a geometry–drive competition. (ii) For
cylindrical pores ($\alpha=0^\circ$), $\langle\tau\rangle$ decreases
with $L_p$ because a longer pore hosts more monomers under a uniform
drive. (iii) Finite-size scaling collapses of $\langle\tau\rangle$
versus $N$ and $L_p$ are obtained with $\langle\tau\rangle\sim
N^{\gamma}F(L_p N^{\phi})$; the best collapses give $\gamma\simeq 3.0$
and $\phi\simeq 1.5$, with a power-law scaling function $F(x)\sim
x^{-p}$.

The opposite trends of $\langle\tau\rangle(\alpha)$ at small and large
$L_p$ can be understood from two coupled effects. First, the widening
exit reduces the entropic confinement penalty and local wall-mediated
hindrance for the leading part of the chain, which speeds up the final
emptying for short pores (entropic relief). Second, in a cone the local
field decreases along the pore, $f_{\mathrm{ext}}(x)=f_0
d/(d+2x\tan\alpha)$ (Eq.~\eqref{eq:fext}), so an extended weak-drive
region develops near the exit; for long pores this slow final emptying
dominates and can reverse the trend, making $\langle\tau\rangle$
increase with $\alpha$. We emphasize that we use this picture only
qualitatively and rely on the scaling collapses for robust quantitative
statements.

Practically, these results identify two independent design levers for
nanopore transport: (i) increasing $L_p$ (at $\alpha=0^\circ$) boosts
the number of driven monomers and accelerates translocation; (ii) for
conical pores, small $\alpha$ values can improve throughput via entropic
assistance, but beyond a geometry-dependent crossover the weak-field
exit becomes a bottleneck. Pore patterning further modulates residence
near the exit: a repulsive exit expedites release, while attractive
exits prolong it, suggesting protocols for sequence-sensitive
discrimination based on the full $(L_p,\alpha)$ space.

Our $\langle\tau\rangle(\alpha)$ crossover is consistent with
cone-induced entropic driving reported by Fazli and
co-workers~\cite{nikoofard2013directed,nikoofard2015flexible}, where a
blob-based confinement-entropy picture predicts a non-monotonic entropic
bias with $\alpha$ and a saturation for wide exits. In our driven
setting, the same entropic relief operates at short $L_p$ (decrease and
saturation with $\alpha$), while for large $L_p$ the position-dependent
external field in the cone dominates and can reverse the trend (increase
with $\alpha$). For $\alpha=0^\circ$, our observation that longer pores
reduce $\langle\tau\rangle$ aligns with earlier findings that extended,
fully driven segments transit faster than short pores under uniform
fields.

The pair $(\gamma,\phi)\approx(3.0,1.5)$ provides high-quality collapses
across drives. The exponent $p$ that characterizes $F(x)\sim x^{-p}$ is
found to be of order unity; \textit{importantly, its precise value may
depend on geometry and patterning}. In particular, collapses with
$p\simeq 0.8$ (cylindrical patterned pores) and $p\simeq 0.6$ (conical
patterned pores) also provide good descriptions (Figs.~6–7). \textit{To
avoid over-claiming universality}, we therefore report that $p$ is
\emph{geometry-sensitive} in our finite ranges of $(L_p,\alpha)$,
whereas $(\gamma,\phi)$ are robust within uncertainties. The measured
$\beta$ from $\langle\tau\rangle\sim N^{\beta}$ is consistent, within
error, with the relation $\beta\approx \gamma-(\phi p)/2$ obtained from
the scaling form.

Our study is two-dimensional and neglects hydrodynamic interactions. Furthermore, we
employ reduced units and a specific choice of attractive/repulsive
patterns. Entrance beads are attractive by design to ensure capture.
Chain lengths and pore lengths are finite and selected to balance
resolution with sampling (1500–2000 successful events per point). These
choices may shift the apparent crossover in $\langle\tau\rangle(\alpha)$
and the fitted $p$ relative to other models or experiments.

Extending the present analysis to 3D with hydrodynamic interactions,
exploring broader $(N,L_p,\alpha)$ ranges, and directly measuring
spatially resolved observables (e.g., monomer waiting-time/residence
profiles along the pore axis and the distribution of the final emptying
time) would sharpen the geometry--drive disentanglement. It would also
be instructive to vary the entrance capture condition (non-attractive
entrance) and to test other patterning ratios to map out $p(\alpha)$
more completely. Finally, comparing with experiments that can tune cone
angles and patterning independently would help assess how the crossover
scale depends on solvent, voltage, and surface chemistry.

\medskip

\section{Acknowledgement}

The authors acknowledge the IISER Mohali computing facility for providing the
necessary resources and support for this work.


\begin{thebibliography}{43}%
\makeatletter
\providecommand \@ifxundefined [1]{%
 \@ifx{#1\undefined}
}%
\providecommand \@ifnum [1]{%
 \ifnum #1\expandafter \@firstoftwo
 \else \expandafter \@secondoftwo
 \fi
}%
\providecommand \@ifx [1]{%
 \ifx #1\expandafter \@firstoftwo
 \else \expandafter \@secondoftwo
 \fi
}%
\providecommand \natexlab [1]{#1}%
\providecommand \enquote  [1]{``#1''}%
\providecommand \bibnamefont  [1]{#1}%
\providecommand \bibfnamefont [1]{#1}%
\providecommand \citenamefont [1]{#1}%
\providecommand \href@noop [0]{\@secondoftwo}%
\providecommand \href [0]{\begingroup \@sanitize@url \@href}%
\providecommand \@href[1]{\@@startlink{#1}\@@href}%
\providecommand \@@href[1]{\endgroup#1\@@endlink}%
\providecommand \@sanitize@url [0]{\catcode `\\12\catcode `\$12\catcode
  `\&12\catcode `\#12\catcode `\^12\catcode `\_12\catcode `\%12\relax}%
\providecommand \@@startlink[1]{}%
\providecommand \@@endlink[0]{}%
\providecommand \url  [0]{\begingroup\@sanitize@url \@url }%
\providecommand \@url [1]{\endgroup\@href {#1}{\urlprefix }}%
\providecommand \urlprefix  [0]{URL }%
\providecommand \Eprint [0]{\href }%
\providecommand \doibase [0]{http://dx.doi.org/}%
\providecommand \selectlanguage [0]{\@gobble}%
\providecommand \bibinfo  [0]{\@secondoftwo}%
\providecommand \bibfield  [0]{\@secondoftwo}%
\providecommand \translation [1]{[#1]}%
\providecommand \BibitemOpen [0]{}%
\providecommand \bibitemStop [0]{}%
\providecommand \bibitemNoStop [0]{.\EOS\space}%
\providecommand \EOS [0]{\spacefactor3000\relax}%
\providecommand \BibitemShut  [1]{\csname bibitem#1\endcsname}%
\let\auto@bib@innerbib\@empty
%</preamble>
\bibitem [{\citenamefont {Alberts}\ \emph {et~al.}(2002)\citenamefont
  {Alberts}, \citenamefont {Bray}, \citenamefont {Lewis}, \citenamefont {Raff},
  \citenamefont {Roberts},\ and\ \citenamefont {Watson}}]{Albe_2002_book}%
  \BibitemOpen
  \bibfield  {author} {\bibinfo {author} {\bibfnamefont {B.}~\bibnamefont
  {Alberts}}, \bibinfo {author} {\bibfnamefont {D.}~\bibnamefont {Bray}},
  \bibinfo {author} {\bibfnamefont {J.}~\bibnamefont {Lewis}}, \bibinfo
  {author} {\bibfnamefont {M.}~\bibnamefont {Raff}}, \bibinfo {author}
  {\bibfnamefont {K.}~\bibnamefont {Roberts}}, \ and\ \bibinfo {author}
  {\bibfnamefont {J.}~\bibnamefont {Watson}},\ }\href@noop {} {\emph {\bibinfo
  {title} {{Molecular Biology of the Cell}}}},\ \bibinfo {edition} {4th}\ ed.\
  (\bibinfo  {publisher} {Garland},\ \bibinfo {year} {2002})\BibitemShut
  {NoStop}%
\bibitem [{\citenamefont {van~der Laan}\ \emph {et~al.}(2007)\citenamefont
  {van~der Laan}, \citenamefont {Meinecke}, \citenamefont {Dudek},
  \citenamefont {Hutu}, \citenamefont {Lind}, \citenamefont {Perschil},
  \citenamefont {Guiard}, \citenamefont {Wagner}, \citenamefont {Pfanner},\
  and\ \citenamefont {Rehling}}]{van2007motor}%
  \BibitemOpen
  \bibfield  {author} {\bibinfo {author} {\bibfnamefont {M.}~\bibnamefont
  {van~der Laan}}, \bibinfo {author} {\bibfnamefont {M.}~\bibnamefont
  {Meinecke}}, \bibinfo {author} {\bibfnamefont {J.}~\bibnamefont {Dudek}},
  \bibinfo {author} {\bibfnamefont {D.~P.}\ \bibnamefont {Hutu}}, \bibinfo
  {author} {\bibfnamefont {M.}~\bibnamefont {Lind}}, \bibinfo {author}
  {\bibfnamefont {I.}~\bibnamefont {Perschil}}, \bibinfo {author}
  {\bibfnamefont {B.}~\bibnamefont {Guiard}}, \bibinfo {author} {\bibfnamefont
  {R.}~\bibnamefont {Wagner}}, \bibinfo {author} {\bibfnamefont
  {N.}~\bibnamefont {Pfanner}}, \ and\ \bibinfo {author} {\bibfnamefont
  {P.}~\bibnamefont {Rehling}},\ }\href@noop {} {\bibfield  {journal} {\bibinfo
   {journal} {Nature cell biology}\ }\textbf {\bibinfo {volume} {9}},\ \bibinfo
  {pages} {1152} (\bibinfo {year} {2007})}\BibitemShut {NoStop}%
\bibitem [{\citenamefont {Salman}\ \emph {et~al.}(2001)\citenamefont {Salman},
  \citenamefont {Zbaida}, \citenamefont {Rabin}, \citenamefont {Chatenay},\
  and\ \citenamefont {Elbaum}}]{salman2001pnas}%
  \BibitemOpen
  \bibfield  {author} {\bibinfo {author} {\bibfnamefont {H.}~\bibnamefont
  {Salman}}, \bibinfo {author} {\bibfnamefont {D.}~\bibnamefont {Zbaida}},
  \bibinfo {author} {\bibfnamefont {Y.}~\bibnamefont {Rabin}}, \bibinfo
  {author} {\bibfnamefont {D.}~\bibnamefont {Chatenay}}, \ and\ \bibinfo
  {author} {\bibfnamefont {M.}~\bibnamefont {Elbaum}},\ }\href {\doibase
  10.1073/pnas.121067698} {\bibfield  {journal} {\bibinfo  {journal}
  {Proceedings of the National Academy of Sciences}\ }\textbf {\bibinfo
  {volume} {98}},\ \bibinfo {pages} {7247} (\bibinfo {year}
  {2001})}\BibitemShut {NoStop}%
\bibitem [{\citenamefont {Ying}\ \emph {et~al.}(2022)\citenamefont {Ying},
  \citenamefont {Hu}, \citenamefont {Zhang}, \citenamefont {Qing},
  \citenamefont {Fragasso}, \citenamefont {Maglia}, \citenamefont {Meller},
  \citenamefont {Bayley}, \citenamefont {Dekker},\ and\ \citenamefont
  {Long}}]{ying2022nn}%
  \BibitemOpen
  \bibfield  {author} {\bibinfo {author} {\bibfnamefont {Y.-L.}\ \bibnamefont
  {Ying}}, \bibinfo {author} {\bibfnamefont {Z.-L.}\ \bibnamefont {Hu}},
  \bibinfo {author} {\bibfnamefont {S.}~\bibnamefont {Zhang}}, \bibinfo
  {author} {\bibfnamefont {Y.}~\bibnamefont {Qing}}, \bibinfo {author}
  {\bibfnamefont {A.}~\bibnamefont {Fragasso}}, \bibinfo {author}
  {\bibfnamefont {G.}~\bibnamefont {Maglia}}, \bibinfo {author} {\bibfnamefont
  {A.}~\bibnamefont {Meller}}, \bibinfo {author} {\bibfnamefont
  {H.}~\bibnamefont {Bayley}}, \bibinfo {author} {\bibfnamefont
  {C.}~\bibnamefont {Dekker}}, \ and\ \bibinfo {author} {\bibfnamefont {Y.-T.}\
  \bibnamefont {Long}},\ }\href {\doibase 10.1038/s41565-022-01193-2}
  {\bibfield  {journal} {\bibinfo  {journal} {Nat. Nanotechnol.}\ }\textbf
  {\bibinfo {volume} {17}},\ \bibinfo {pages} {1136} (\bibinfo {year}
  {2022})}\BibitemShut {NoStop}%
\bibitem [{\citenamefont {Kasianowicz}\ \emph {et~al.}(1996)\citenamefont
  {Kasianowicz}, \citenamefont {Brandin}, \citenamefont {Branton},\ and\
  \citenamefont {Deamer}}]{kasianowicz1996characterization}%
  \BibitemOpen
  \bibfield  {author} {\bibinfo {author} {\bibfnamefont {J.~J.}\ \bibnamefont
  {Kasianowicz}}, \bibinfo {author} {\bibfnamefont {E.}~\bibnamefont
  {Brandin}}, \bibinfo {author} {\bibfnamefont {D.}~\bibnamefont {Branton}}, \
  and\ \bibinfo {author} {\bibfnamefont {D.~W.}\ \bibnamefont {Deamer}},\
  }\href@noop {} {\bibfield  {journal} {\bibinfo  {journal} {Proceedings of the
  National Academy of Sciences}\ }\textbf {\bibinfo {volume} {93}},\ \bibinfo
  {pages} {13770} (\bibinfo {year} {1996})}\BibitemShut {NoStop}%
\bibitem [{\citenamefont {Muthukumar}(2007)}]{Muthukumar2007-yc}%
  \BibitemOpen
  \bibfield  {author} {\bibinfo {author} {\bibfnamefont {M.}~\bibnamefont
  {Muthukumar}},\ }\href@noop {} {\bibfield  {journal} {\bibinfo  {journal}
  {Annu. Rev. Biophys. Biomol. Struct.}\ }\textbf {\bibinfo {volume} {36}},\
  \bibinfo {pages} {435} (\bibinfo {year} {2007})}\BibitemShut {NoStop}%
\bibitem [{\citenamefont {Muthukumar}(2011)}]{muthukumar2}%
  \BibitemOpen
  \bibfield  {author} {\bibinfo {author} {\bibfnamefont {M.}~\bibnamefont
  {Muthukumar}},\ }\href@noop {} {\emph {\bibinfo {title} {Polymer
  Translocation (1st ed.).}}}\ (\bibinfo  {publisher} {CRC Press.},\ \bibinfo
  {year} {2011})\BibitemShut {NoStop}%
\bibitem [{\citenamefont {Sung}\ and\ \citenamefont
  {Park}(1996)}]{sung1996polymer}%
  \BibitemOpen
  \bibfield  {author} {\bibinfo {author} {\bibfnamefont {W.}~\bibnamefont
  {Sung}}\ and\ \bibinfo {author} {\bibfnamefont {P.}~\bibnamefont {Park}},\
  }\href@noop {} {\bibfield  {journal} {\bibinfo  {journal} {Physical review
  letters}\ }\textbf {\bibinfo {volume} {77}},\ \bibinfo {pages} {783}
  (\bibinfo {year} {1996})}\BibitemShut {NoStop}%
\bibitem [{\citenamefont {Palyulin}\ \emph {et~al.}(2014)\citenamefont
  {Palyulin}, \citenamefont {Ala-Nissila},\ and\ \citenamefont
  {Metzler}}]{palyulin2014polymer}%
  \BibitemOpen
  \bibfield  {author} {\bibinfo {author} {\bibfnamefont {V.~V.}\ \bibnamefont
  {Palyulin}}, \bibinfo {author} {\bibfnamefont {T.}~\bibnamefont
  {Ala-Nissila}}, \ and\ \bibinfo {author} {\bibfnamefont {R.}~\bibnamefont
  {Metzler}},\ }\href@noop {} {\bibfield  {journal} {\bibinfo  {journal} {Soft
  matter}\ }\textbf {\bibinfo {volume} {10}},\ \bibinfo {pages} {9016}
  (\bibinfo {year} {2014})}\BibitemShut {NoStop}%
\bibitem [{\citenamefont {Rowghanian}\ and\ \citenamefont
  {Grosberg}(2011)}]{rowghanian2011force}%
  \BibitemOpen
  \bibfield  {author} {\bibinfo {author} {\bibfnamefont {P.}~\bibnamefont
  {Rowghanian}}\ and\ \bibinfo {author} {\bibfnamefont {A.~Y.}\ \bibnamefont
  {Grosberg}},\ }\href@noop {} {\bibfield  {journal} {\bibinfo  {journal} {The
  Journal of Physical Chemistry B}\ }\textbf {\bibinfo {volume} {115}},\
  \bibinfo {pages} {14127} (\bibinfo {year} {2011})}\BibitemShut {NoStop}%
\bibitem [{\citenamefont {Sakaue}(2007)}]{sakaue2007nonequilibrium}%
  \BibitemOpen
  \bibfield  {author} {\bibinfo {author} {\bibfnamefont {T.}~\bibnamefont
  {Sakaue}},\ }\href@noop {} {\bibfield  {journal} {\bibinfo  {journal}
  {Physical Review E}\ }\textbf {\bibinfo {volume} {76}},\ \bibinfo {pages}
  {021803} (\bibinfo {year} {2007})}\BibitemShut {NoStop}%
\bibitem [{\citenamefont {Kantor}\ and\ \citenamefont
  {Kardar}(2004)}]{kantor2004anomalous}%
  \BibitemOpen
  \bibfield  {author} {\bibinfo {author} {\bibfnamefont {Y.}~\bibnamefont
  {Kantor}}\ and\ \bibinfo {author} {\bibfnamefont {M.}~\bibnamefont
  {Kardar}},\ }\href@noop {} {\bibfield  {journal} {\bibinfo  {journal}
  {Physical Review E}\ }\textbf {\bibinfo {volume} {69}},\ \bibinfo {pages}
  {021806} (\bibinfo {year} {2004})}\BibitemShut {NoStop}%
\bibitem [{\citenamefont {Dubbeldam}\ \emph {et~al.}(2007)\citenamefont
  {Dubbeldam}, \citenamefont {Milchev}, \citenamefont {Rostiashvili},\ and\
  \citenamefont {Vilgis}}]{dubbeldam2007driven}%
  \BibitemOpen
  \bibfield  {author} {\bibinfo {author} {\bibfnamefont {J.}~\bibnamefont
  {Dubbeldam}}, \bibinfo {author} {\bibfnamefont {A.}~\bibnamefont {Milchev}},
  \bibinfo {author} {\bibfnamefont {V.}~\bibnamefont {Rostiashvili}}, \ and\
  \bibinfo {author} {\bibfnamefont {T.~A.}\ \bibnamefont {Vilgis}},\
  }\href@noop {} {\bibfield  {journal} {\bibinfo  {journal} {Europhysics
  Letters}\ }\textbf {\bibinfo {volume} {79}},\ \bibinfo {pages} {18002}
  (\bibinfo {year} {2007})}\BibitemShut {NoStop}%
\bibitem [{\citenamefont {Ikonen}\ \emph
  {et~al.}(2012{\natexlab{a}})\citenamefont {Ikonen}, \citenamefont
  {Bhattacharya}, \citenamefont {Ala-Nissila},\ and\ \citenamefont
  {Sung}}]{ikonen2012unifying}%
  \BibitemOpen
  \bibfield  {author} {\bibinfo {author} {\bibfnamefont {T.}~\bibnamefont
  {Ikonen}}, \bibinfo {author} {\bibfnamefont {A.}~\bibnamefont
  {Bhattacharya}}, \bibinfo {author} {\bibfnamefont {T.}~\bibnamefont
  {Ala-Nissila}}, \ and\ \bibinfo {author} {\bibfnamefont {W.}~\bibnamefont
  {Sung}},\ }\href@noop {} {\bibfield  {journal} {\bibinfo  {journal} {Physical
  Review E}\ }\textbf {\bibinfo {volume} {85}},\ \bibinfo {pages} {051803}
  (\bibinfo {year} {2012}{\natexlab{a}})}\BibitemShut {NoStop}%
\bibitem [{\citenamefont {Muthukumar}(1999)}]{muthukumar1999polymer}%
  \BibitemOpen
  \bibfield  {author} {\bibinfo {author} {\bibfnamefont {M.}~\bibnamefont
  {Muthukumar}},\ }\href@noop {} {\bibfield  {journal} {\bibinfo  {journal}
  {The Journal of Chemical Physics}\ }\textbf {\bibinfo {volume} {111}},\
  \bibinfo {pages} {10371} (\bibinfo {year} {1999})}\BibitemShut {NoStop}%
\bibitem [{\citenamefont {Ikonen}\ \emph
  {et~al.}(2012{\natexlab{b}})\citenamefont {Ikonen}, \citenamefont
  {Bhattacharya}, \citenamefont {Ala-Nissila},\ and\ \citenamefont
  {Sung}}]{ikonen2012influence}%
  \BibitemOpen
  \bibfield  {author} {\bibinfo {author} {\bibfnamefont {T.}~\bibnamefont
  {Ikonen}}, \bibinfo {author} {\bibfnamefont {A.}~\bibnamefont
  {Bhattacharya}}, \bibinfo {author} {\bibfnamefont {T.}~\bibnamefont
  {Ala-Nissila}}, \ and\ \bibinfo {author} {\bibfnamefont {W.}~\bibnamefont
  {Sung}},\ }\href@noop {} {\bibfield  {journal} {\bibinfo  {journal} {The
  Journal of chemical physics}\ }\textbf {\bibinfo {volume} {137}},\ \bibinfo
  {pages} {085101} (\bibinfo {year} {2012}{\natexlab{b}})}\BibitemShut
  {NoStop}%
\bibitem [{\citenamefont {Sarabadani}\ \emph {et~al.}(2014)\citenamefont
  {Sarabadani}, \citenamefont {Ikonen},\ and\ \citenamefont
  {Ala-Nissila}}]{sarabadani2014iso}%
  \BibitemOpen
  \bibfield  {author} {\bibinfo {author} {\bibfnamefont {J.}~\bibnamefont
  {Sarabadani}}, \bibinfo {author} {\bibfnamefont {T.}~\bibnamefont {Ikonen}},
  \ and\ \bibinfo {author} {\bibfnamefont {T.}~\bibnamefont {Ala-Nissila}},\
  }\href@noop {} {\bibfield  {journal} {\bibinfo  {journal} {The Journal of
  Chemical Physics}\ }\textbf {\bibinfo {volume} {141}},\ \bibinfo {pages}
  {214907} (\bibinfo {year} {2014})}\BibitemShut {NoStop}%
\bibitem [{\citenamefont {Sarabadani}\ and\ \citenamefont
  {Ala-Nissila}(2018)}]{sarabadani2018theory}%
  \BibitemOpen
  \bibfield  {author} {\bibinfo {author} {\bibfnamefont {J.}~\bibnamefont
  {Sarabadani}}\ and\ \bibinfo {author} {\bibfnamefont {T.}~\bibnamefont
  {Ala-Nissila}},\ }\href@noop {} {\bibfield  {journal} {\bibinfo  {journal}
  {Journal of Physics: Condensed Matter}\ }\textbf {\bibinfo {volume} {30}},\
  \bibinfo {pages} {274002} (\bibinfo {year} {2018})}\BibitemShut {NoStop}%
\bibitem [{\citenamefont {Dubbeldam}\ \emph {et~al.}(2012)\citenamefont
  {Dubbeldam}, \citenamefont {Rostiashvili}, \citenamefont {Milchev},\ and\
  \citenamefont {Vilgis}}]{dubbeldamPRE2012}%
  \BibitemOpen
  \bibfield  {author} {\bibinfo {author} {\bibfnamefont {J.~L.~A.}\
  \bibnamefont {Dubbeldam}}, \bibinfo {author} {\bibfnamefont {V.~G.}\
  \bibnamefont {Rostiashvili}}, \bibinfo {author} {\bibfnamefont
  {A.}~\bibnamefont {Milchev}}, \ and\ \bibinfo {author} {\bibfnamefont
  {T.~A.}\ \bibnamefont {Vilgis}},\ }\href {\doibase
  10.1103/PhysRevE.85.041801} {\bibfield  {journal} {\bibinfo  {journal} {Phys.
  Rev. E}\ }\textbf {\bibinfo {volume} {85}},\ \bibinfo {pages} {041801}
  (\bibinfo {year} {2012})}\BibitemShut {NoStop}%
\bibitem [{\citenamefont {Chen}\ \emph {et~al.}(2021)\citenamefont {Chen},
  \citenamefont {Jou}, \citenamefont {Ermann}, \citenamefont {Muthukumar},
  \citenamefont {Keyser},\ and\ \citenamefont {Bell}}]{chen2021dynamics}%
  \BibitemOpen
  \bibfield  {author} {\bibinfo {author} {\bibfnamefont {K.}~\bibnamefont
  {Chen}}, \bibinfo {author} {\bibfnamefont {I.}~\bibnamefont {Jou}}, \bibinfo
  {author} {\bibfnamefont {N.}~\bibnamefont {Ermann}}, \bibinfo {author}
  {\bibfnamefont {M.}~\bibnamefont {Muthukumar}}, \bibinfo {author}
  {\bibfnamefont {U.~F.}\ \bibnamefont {Keyser}}, \ and\ \bibinfo {author}
  {\bibfnamefont {N.~A.}\ \bibnamefont {Bell}},\ }\href@noop {} {\bibfield
  {journal} {\bibinfo  {journal} {Nature Physics}\ }\textbf {\bibinfo {volume}
  {17}},\ \bibinfo {pages} {1043} (\bibinfo {year} {2021})}\BibitemShut
  {NoStop}%
\bibitem [{\citenamefont {Milchev}(2011)}]{milchev2011single}%
  \BibitemOpen
  \bibfield  {author} {\bibinfo {author} {\bibfnamefont {A.}~\bibnamefont
  {Milchev}},\ }\href@noop {} {\bibfield  {journal} {\bibinfo  {journal}
  {Journal of Physics: Condensed Matter}\ }\textbf {\bibinfo {volume} {23}},\
  \bibinfo {pages} {103101} (\bibinfo {year} {2011})}\BibitemShut {NoStop}%
\bibitem [{\citenamefont {Chuang}\ \emph {et~al.}(2001)\citenamefont {Chuang},
  \citenamefont {Kantor},\ and\ \citenamefont {Kardar}}]{chuang2001anomalous}%
  \BibitemOpen
  \bibfield  {author} {\bibinfo {author} {\bibfnamefont {J.}~\bibnamefont
  {Chuang}}, \bibinfo {author} {\bibfnamefont {Y.}~\bibnamefont {Kantor}}, \
  and\ \bibinfo {author} {\bibfnamefont {M.}~\bibnamefont {Kardar}},\
  }\href@noop {} {\bibfield  {journal} {\bibinfo  {journal} {Physical Review
  E}\ }\textbf {\bibinfo {volume} {65}},\ \bibinfo {pages} {011802} (\bibinfo
  {year} {2001})}\BibitemShut {NoStop}%
\bibitem [{\citenamefont {Luo}\ \emph {et~al.}(2007)\citenamefont {Luo},
  \citenamefont {Ala-Nissila}, \citenamefont {Ying},\ and\ \citenamefont
  {Bhattacharya}}]{luo2007influence}%
  \BibitemOpen
  \bibfield  {author} {\bibinfo {author} {\bibfnamefont {K.}~\bibnamefont
  {Luo}}, \bibinfo {author} {\bibfnamefont {T.}~\bibnamefont {Ala-Nissila}},
  \bibinfo {author} {\bibfnamefont {S.-C.}\ \bibnamefont {Ying}}, \ and\
  \bibinfo {author} {\bibfnamefont {A.}~\bibnamefont {Bhattacharya}},\
  }\href@noop {} {\bibfield  {journal} {\bibinfo  {journal} {Physical Review
  Letters}\ }\textbf {\bibinfo {volume} {99}},\ \bibinfo {pages} {148102}
  (\bibinfo {year} {2007})}\BibitemShut {NoStop}%
\bibitem [{\citenamefont {Doi}\ \emph {et~al.}(1988)\citenamefont {Doi},
  \citenamefont {Edwards},\ and\ \citenamefont {Edwards}}]{doi1988theory}%
  \BibitemOpen
  \bibfield  {author} {\bibinfo {author} {\bibfnamefont {M.}~\bibnamefont
  {Doi}}, \bibinfo {author} {\bibfnamefont {S.~F.}\ \bibnamefont {Edwards}}, \
  and\ \bibinfo {author} {\bibfnamefont {S.~F.}\ \bibnamefont {Edwards}},\
  }\href@noop {} {\emph {\bibinfo {title} {The theory of polymer dynamics}}},\
  Vol.~\bibinfo {volume} {73}\ (\bibinfo  {publisher} {oxford university
  press},\ \bibinfo {year} {1988})\BibitemShut {NoStop}%
\bibitem [{\citenamefont {Storm}\ \emph {et~al.}(2003)\citenamefont {Storm},
  \citenamefont {Chen}, \citenamefont {Ling}, \citenamefont {Zandbergen},\ and\
  \citenamefont {Dekker}}]{storm2003fabrication}%
  \BibitemOpen
  \bibfield  {author} {\bibinfo {author} {\bibfnamefont {A.}~\bibnamefont
  {Storm}}, \bibinfo {author} {\bibfnamefont {J.}~\bibnamefont {Chen}},
  \bibinfo {author} {\bibfnamefont {X.}~\bibnamefont {Ling}}, \bibinfo {author}
  {\bibfnamefont {H.}~\bibnamefont {Zandbergen}}, \ and\ \bibinfo {author}
  {\bibfnamefont {C.}~\bibnamefont {Dekker}},\ }\href@noop {} {\bibfield
  {journal} {\bibinfo  {journal} {Nature materials}\ }\textbf {\bibinfo
  {volume} {2}},\ \bibinfo {pages} {537} (\bibinfo {year} {2003})}\BibitemShut
  {NoStop}%
\bibitem [{\citenamefont {Wanunu}\ and\ \citenamefont
  {Meller}(2007)}]{wanunu2007chemically}%
  \BibitemOpen
  \bibfield  {author} {\bibinfo {author} {\bibfnamefont {M.}~\bibnamefont
  {Wanunu}}\ and\ \bibinfo {author} {\bibfnamefont {A.}~\bibnamefont
  {Meller}},\ }\href@noop {} {\bibfield  {journal} {\bibinfo  {journal} {Nano
  letters}\ }\textbf {\bibinfo {volume} {7}},\ \bibinfo {pages} {1580}
  (\bibinfo {year} {2007})}\BibitemShut {NoStop}%
\bibitem [{\citenamefont {Chen}\ \emph {et~al.}(2004)\citenamefont {Chen},
  \citenamefont {Gu}, \citenamefont {Brandin}, \citenamefont {Kim},
  \citenamefont {Wang},\ and\ \citenamefont {Branton}}]{chen2004probing}%
  \BibitemOpen
  \bibfield  {author} {\bibinfo {author} {\bibfnamefont {P.}~\bibnamefont
  {Chen}}, \bibinfo {author} {\bibfnamefont {J.}~\bibnamefont {Gu}}, \bibinfo
  {author} {\bibfnamefont {E.}~\bibnamefont {Brandin}}, \bibinfo {author}
  {\bibfnamefont {Y.-R.}\ \bibnamefont {Kim}}, \bibinfo {author} {\bibfnamefont
  {Q.}~\bibnamefont {Wang}}, \ and\ \bibinfo {author} {\bibfnamefont
  {D.}~\bibnamefont {Branton}},\ }\href@noop {} {\bibfield  {journal} {\bibinfo
   {journal} {Nano letters}\ }\textbf {\bibinfo {volume} {4}},\ \bibinfo
  {pages} {2293} (\bibinfo {year} {2004})}\BibitemShut {NoStop}%
\bibitem [{\citenamefont {Lan}\ \emph {et~al.}(2011)\citenamefont {Lan},
  \citenamefont {Holden},\ and\ \citenamefont {White}}]{lan2011pressure}%
  \BibitemOpen
  \bibfield  {author} {\bibinfo {author} {\bibfnamefont {W.-J.}\ \bibnamefont
  {Lan}}, \bibinfo {author} {\bibfnamefont {D.~A.}\ \bibnamefont {Holden}}, \
  and\ \bibinfo {author} {\bibfnamefont {H.~S.}\ \bibnamefont {White}},\
  }\href@noop {} {\bibfield  {journal} {\bibinfo  {journal} {Journal of the
  American Chemical Society}\ }\textbf {\bibinfo {volume} {133}},\ \bibinfo
  {pages} {13300} (\bibinfo {year} {2011})}\BibitemShut {NoStop}%
\bibitem [{\citenamefont {Liu}\ \emph {et~al.}(2012)\citenamefont {Liu},
  \citenamefont {Wu}, \citenamefont {Wu}, \citenamefont {Xie}, \citenamefont
  {Kong}, \citenamefont {Ye},\ and\ \citenamefont {Liu}}]{liu2012voltage}%
  \BibitemOpen
  \bibfield  {author} {\bibinfo {author} {\bibfnamefont {Q.}~\bibnamefont
  {Liu}}, \bibinfo {author} {\bibfnamefont {H.}~\bibnamefont {Wu}}, \bibinfo
  {author} {\bibfnamefont {L.}~\bibnamefont {Wu}}, \bibinfo {author}
  {\bibfnamefont {X.}~\bibnamefont {Xie}}, \bibinfo {author} {\bibfnamefont
  {J.}~\bibnamefont {Kong}}, \bibinfo {author} {\bibfnamefont {X.}~\bibnamefont
  {Ye}}, \ and\ \bibinfo {author} {\bibfnamefont {L.}~\bibnamefont {Liu}},\
  }\href@noop {} {\bibfield  {journal} {\bibinfo  {journal} {PLoS one}\ } {\bibinfo {volume} {7}},\ \bibinfo {pages} {e46014}
  (\bibinfo {year} {2012})}\BibitemShut {NoStop}%
\bibitem [{\citenamefont {Luo}\ \emph {et~al.}(2008)\citenamefont {Luo},
  \citenamefont {Ala-Nissila}, \citenamefont {Ying},\ and\ \citenamefont
  {Bhattacharya}}]{luo2008translocation}%
  \BibitemOpen
  \bibfield  {author} {\bibinfo {author} {\bibfnamefont {K.}~\bibnamefont
  {Luo}}, \bibinfo {author} {\bibfnamefont {T.}~\bibnamefont {Ala-Nissila}},
  \bibinfo {author} {\bibfnamefont {S.-C.}\ \bibnamefont {Ying}}, \ and\
  \bibinfo {author} {\bibfnamefont {A.}~\bibnamefont {Bhattacharya}},\
  }\href@noop {} {\bibfield  {journal} {\bibinfo  {journal} {Physical Review
  E}\ }\textbf {\bibinfo {volume} {78}},\ \bibinfo {pages} {061918} (\bibinfo
  {year} {2008})}\BibitemShut {NoStop}%
\bibitem [{\citenamefont {Cohen}\ \emph
  {et~al.}(2012{\natexlab{a}})\citenamefont {Cohen}, \citenamefont
  {Chaudhuri},\ and\ \citenamefont {Golestanian}}]{cohen2012stochastic}%
  \BibitemOpen
  \bibfield  {author} {\bibinfo {author} {\bibfnamefont {J.~A.}\ \bibnamefont
  {Cohen}}, \bibinfo {author} {\bibfnamefont {A.}~\bibnamefont {Chaudhuri}}, \
  and\ \bibinfo {author} {\bibfnamefont {R.}~\bibnamefont {Golestanian}},\
  }\href@noop {} {\bibfield  {journal} {\bibinfo  {journal} {Physical Review
  X}\ }\textbf {\bibinfo {volume} {2}},\ \bibinfo {pages} {021002} (\bibinfo
  {year} {2012}{\natexlab{a}})}\BibitemShut {NoStop}%
\bibitem [{\citenamefont {Cohen}\ \emph
  {et~al.}(2012{\natexlab{b}})\citenamefont {Cohen}, \citenamefont
  {Chaudhuri},\ and\ \citenamefont {Golestanian}}]{cohen2012translocation}%
  \BibitemOpen
  \bibfield  {author} {\bibinfo {author} {\bibfnamefont {J.~A.}\ \bibnamefont
  {Cohen}}, \bibinfo {author} {\bibfnamefont {A.}~\bibnamefont {Chaudhuri}}, \
  and\ \bibinfo {author} {\bibfnamefont {R.}~\bibnamefont {Golestanian}},\
  }\href@noop {} {\bibfield  {journal} {\bibinfo  {journal} {The Journal of
  chemical physics}\ }\textbf {\bibinfo {volume} {137}},\ \bibinfo {pages}
  {204911} (\bibinfo {year} {2012}{\natexlab{b}})}\BibitemShut {NoStop}%
\bibitem [{\citenamefont {Katkar}\ and\ \citenamefont
  {Muthukumar}(2018)}]{katkar2018role}%
  \BibitemOpen
  \bibfield  {author} {\bibinfo {author} {\bibfnamefont {H.~H.}\ \bibnamefont
  {Katkar}}\ and\ \bibinfo {author} {\bibfnamefont {M.}~\bibnamefont
  {Muthukumar}},\ }\href@noop {} {\bibfield  {journal} {\bibinfo  {journal}
  {The Journal of chemical physics}\ }\textbf {\bibinfo {volume} {148}},\
  \bibinfo {pages} {024903} (\bibinfo {year} {2018})}\BibitemShut {NoStop}%
\bibitem [{\citenamefont {Kumar}\ \emph {et~al.}(2018)\citenamefont {Kumar},
  \citenamefont {Chaudhuri},\ and\ \citenamefont
  {Kapri}}]{kumar2018sequencing}%
  \BibitemOpen
  \bibfield  {author} {\bibinfo {author} {\bibfnamefont {R.}~\bibnamefont
  {Kumar}}, \bibinfo {author} {\bibfnamefont {A.}~\bibnamefont {Chaudhuri}}, \
  and\ \bibinfo {author} {\bibfnamefont {R.}~\bibnamefont {Kapri}},\
  }\href@noop {} {\bibfield  {journal} {\bibinfo  {journal} {The Journal of
  chemical physics}\ }\textbf {\bibinfo {volume} {148}},\ \bibinfo {pages}
  {164901} (\bibinfo {year} {2018})}\BibitemShut {NoStop}%
\bibitem [{\citenamefont {Nikoofard}\ \emph {et~al.}(2013)\citenamefont
  {Nikoofard}, \citenamefont {Khalilian},\ and\ \citenamefont
  {Fazli}}]{nikoofard2013directed}%
  \BibitemOpen
  \bibfield  {author} {\bibinfo {author} {\bibfnamefont {N.}~\bibnamefont
  {Nikoofard}}, \bibinfo {author} {\bibfnamefont {H.}~\bibnamefont
  {Khalilian}}, \ and\ \bibinfo {author} {\bibfnamefont {H.}~\bibnamefont
  {Fazli}},\ }\href@noop {} {\bibfield  {journal} {\bibinfo  {journal} {The
  Journal of Chemical Physics}\ }\textbf {\bibinfo {volume} {139}},\ \bibinfo
  {pages} {074901} (\bibinfo {year} {2013})}\BibitemShut {NoStop}%
\bibitem [{\citenamefont {Nikoofard}\ \emph {et~al.}(2014)\citenamefont
  {Nikoofard}, \citenamefont {Hoseinpoor},\ and\ \citenamefont
  {Zahedifar}}]{nikoofard2014accuracy}%
  \BibitemOpen
  \bibfield  {author} {\bibinfo {author} {\bibfnamefont {N.}~\bibnamefont
  {Nikoofard}}, \bibinfo {author} {\bibfnamefont {S.~M.}\ \bibnamefont
  {Hoseinpoor}}, \ and\ \bibinfo {author} {\bibfnamefont {M.}~\bibnamefont
  {Zahedifar}},\ }\href@noop {} {\bibfield  {journal} {\bibinfo  {journal}
  {Physical Review E}\ }\textbf {\bibinfo {volume} {90}},\ \bibinfo {pages}
  {062603} (\bibinfo {year} {2014})}\BibitemShut {NoStop}%
\bibitem [{\citenamefont {Nikoofard}\ and\ \citenamefont
  {Fazli}(2015)}]{nikoofard2015flexible}%
  \BibitemOpen
  \bibfield  {author} {\bibinfo {author} {\bibfnamefont {N.}~\bibnamefont
  {Nikoofard}}\ and\ \bibinfo {author} {\bibfnamefont {H.}~\bibnamefont
  {Fazli}},\ }\href@noop {} {\bibfield  {journal} {\bibinfo  {journal} {Soft
  Matter}\ }\textbf {\bibinfo {volume} {11}},\ \bibinfo {pages} {4879}
  (\bibinfo {year} {2015})}\BibitemShut {NoStop}%
\bibitem [{\citenamefont {Sharma}\ \emph {et~al.}(2022)\citenamefont {Sharma},
  \citenamefont {Kapri},\ and\ \citenamefont {Chaudhuri}}]{sharma2022driven}%
  \BibitemOpen
  \bibfield  {author} {\bibinfo {author} {\bibfnamefont {A.}~\bibnamefont
  {Sharma}}, \bibinfo {author} {\bibfnamefont {R.}~\bibnamefont {Kapri}}, \
  and\ \bibinfo {author} {\bibfnamefont {A.}~\bibnamefont {Chaudhuri}},\
  }\href@noop {} {\bibfield  {journal} {\bibinfo  {journal} {Scientific
  Reports}\ }\textbf {\bibinfo {volume} {12}},\ \bibinfo {pages} {19081}
  (\bibinfo {year} {2022})}\BibitemShut {NoStop}%
\bibitem [{\citenamefont {Sharma}(2024)}]{Sharma_2024}%
  \BibitemOpen
  \bibfield  {author} {\bibinfo {author} {\bibfnamefont {A.}~\bibnamefont
  {Sharma}},\ }\href {\doibase 10.1088/1402-4896/ad3a47} {\bibfield  {journal}
  {\bibinfo  {journal} {Physica Scripta}\ }\textbf {\bibinfo {volume} {99}},\
  \bibinfo {pages} {055025} (\bibinfo {year} {2024})}\BibitemShut {NoStop}%
\bibitem [{\citenamefont {Upadhyay}\ \emph
  {et~al.}(2024{\natexlab{a}})\citenamefont {Upadhyay}, \citenamefont {Kapri},\
  and\ \citenamefont {Chaudhuri}}]{upadhyay2024epje}%
  \BibitemOpen
  \bibfield  {author} {\bibinfo {author} {\bibfnamefont {G.}~\bibnamefont
  {Upadhyay}}, \bibinfo {author} {\bibfnamefont {R.}~\bibnamefont {Kapri}}, \
  and\ \bibinfo {author} {\bibfnamefont {A.}~\bibnamefont {Chaudhuri}},\
  }\href@noop {} {\bibfield  {journal} {\bibinfo  {journal} {Eur. Phys. J. E}\
  }\textbf {\bibinfo {volume} {47}},\ \bibinfo {pages} {23} (\bibinfo {year}
  {2024}{\natexlab{a}})}\BibitemShut {NoStop}%
\bibitem [{\citenamefont {Upadhyay}\ \emph
  {et~al.}(2024{\natexlab{b}})\citenamefont {Upadhyay}, \citenamefont {Kapri},\
  and\ \citenamefont {Chaudhuri}}]{upadhyay2024jpcm}%
  \BibitemOpen
  \bibfield  {author} {\bibinfo {author} {\bibfnamefont {G.}~\bibnamefont
  {Upadhyay}}, \bibinfo {author} {\bibfnamefont {R.}~\bibnamefont {Kapri}}, \
  and\ \bibinfo {author} {\bibfnamefont {A.}~\bibnamefont {Chaudhuri}},\ }\href
  {\doibase 10.1088/1361-648X/ad21a9} {\bibfield  {journal} {\bibinfo
  {journal} {J. Phys.: Condens. Matter}\ }\textbf {\bibinfo {volume} {36}},\
  \bibinfo {pages} {185101} (\bibinfo {year} {2024}{\natexlab{b}})}\BibitemShut
  {NoStop}%
\bibitem [{\citenamefont {Grest}\ and\ \citenamefont
  {Kremer}(1986)}]{grest1986molecular}%
  \BibitemOpen
  \bibfield  {author} {\bibinfo {author} {\bibfnamefont {G.~S.}\ \bibnamefont
  {Grest}}\ and\ \bibinfo {author} {\bibfnamefont {K.}~\bibnamefont {Kremer}},\
  }\href@noop {} {\bibfield  {journal} {\bibinfo  {journal} {Physical Review
  A}\ }\textbf {\bibinfo {volume} {33}},\ \bibinfo {pages} {3628} (\bibinfo
  {year} {1986})}\BibitemShut {NoStop}%
\bibitem [{\citenamefont {Thompson}\ \emph {et~al.}(2022)\citenamefont
  {Thompson}, \citenamefont {Aktulga}, \citenamefont {Berger}, \citenamefont
  {Bolintineanu}, \citenamefont {Brown}, \citenamefont {Crozier}, \citenamefont
  {in't Veld}, \citenamefont {Kohlmeyer}, \citenamefont {Moore}, \citenamefont
  {Nguyen} \emph {et~al.}}]{thompson2022lammps}%
  \BibitemOpen
  \bibfield  {author} {\bibinfo {author} {\bibfnamefont {A.~P.}\ \bibnamefont
  {Thompson}}, \bibinfo {author} {\bibfnamefont {H.~M.}\ \bibnamefont
  {Aktulga}}, \bibinfo {author} {\bibfnamefont {R.}~\bibnamefont {Berger}},
  \bibinfo {author} {\bibfnamefont {D.~S.}\ \bibnamefont {Bolintineanu}},
  \bibinfo {author} {\bibfnamefont {W.~M.}\ \bibnamefont {Brown}}, \bibinfo
  {author} {\bibfnamefont {P.~S.}\ \bibnamefont {Crozier}}, \bibinfo {author}
  {\bibfnamefont {P.~J.}\ \bibnamefont {in't Veld}}, \bibinfo {author}
  {\bibfnamefont {A.}~\bibnamefont {Kohlmeyer}}, \bibinfo {author}
  {\bibfnamefont {S.~G.}\ \bibnamefont {Moore}}, \bibinfo {author}
  {\bibfnamefont {T.~D.}\ \bibnamefont {Nguyen}},  \emph {et~al.},\ }\href@noop
  {} {\bibfield  {journal} {\bibinfo  {journal} {Computer Physics
  Communications}\ }\textbf {\bibinfo {volume} {271}},\ \bibinfo {pages}
  {108171} (\bibinfo {year} {2022})}\BibitemShut {NoStop}%
\end{thebibliography}
\end{document}